\documentclass[aps,prl,preprintnumbers,superscriptaddress,nofootinbib,twocolumn]{revtex4-2}
\usepackage{graphicx}
\usepackage{bm,latexsym,amsmath,amssymb,amsfonts,mathrsfs}
\usepackage{color}
\input{colordvi.tex}
\usepackage{dirtytalk}
\usepackage{mathtools,mathbbol}
\usepackage{graphicx,hyperref,xcolor}
\usepackage{lipsum}
\usepackage{float}

\newcommand{\LF}{\left(}
\newcommand{\RF}{\right)}

\def\calMa{{V_{4}}}

\def\ba{\begin{array}}
\def\ea{\end{array}} 
\def\bea{\begin{eqnarray}}
\def\eea{\end{eqnarray}}
\def\beq{\begin{equation}}
\def\eeq{\end{equation}}
\def\ben{\begin{enumerate}}
\def\een{\end{enumerate}}

\def\brr{\begin{array}}
\def\err{\end{array}}

\def\calMa{{V_{4}}}

\usepackage[normalem]{ulem}

\begin{document}



\title{Gravitational Bounce from the Quantum Exclusion Principle}

\author{Enrique Gazta\~naga}
\affiliation{Institute of Cosmology \& Gravitation,
	University of Portsmouth,
	Dennis Sciama Building, Burnaby Road,
	Portsmouth, PO1 3FX, United Kingdom}
\affiliation{Institute of Space Sciences (ICE, CSIC), 08193 Barcelona, Spain}
\affiliation{Institut d\'~Estudis Espacials de Catalunya (IEEC), 08034 Barcelona, Spain}

\author{K. Sravan Kumar}
\affiliation{Institute of Cosmology \& Gravitation,
	University of Portsmouth,
	Dennis Sciama Building, Burnaby Road,
	Portsmouth, PO1 3FX, United Kingdom}

\author{S. Pradhan}
\affiliation{Institute of Cosmology \& Gravitation,
	University of Portsmouth,
	Dennis Sciama Building, Burnaby Road,
	Portsmouth, PO1 3FX, United Kingdom}
\affiliation{Department of Physical Sciences, Indian Institute of Science Education and Research (IISER) Kolkata, Mohanpur 741246, India}

\author{M. Gabler}
\affiliation{Departmento de Astronomía y Astrofísica, Universitat de València, E-46100 Burjassot (València), Spain}

\begin{abstract}

We investigate the fully relativistic spherical collapse model of a uniform distribution of mass $M$ with initial comoving radius $\chi_*$ and spatial curvature $k \equiv 1/\chi_k^2 \le 1/\chi_*^2$ representing an over-density or bounded perturbation within a larger background. 
Our model incorporates a perfect fluid with an evolving equation of state, $P = P(\rho)$, which asymptotically transitions from pressureless dust ($P = 0$) to a ground state characterized by a uniform, time-independent energy density $\rho_{\rm G}$. This transition is motivated by the quantum exclusion principle, which prevents singular collapse, as observed in supernova core-collapse explosions. We analytically demonstrate that this transition induces a gravitational bounce at a radius $R_{\rm B} = (8 \pi G \rho_{\rm G}/3)^{-1/2}$. The bounce leads to an exponential expansion phase, where $P(\rho)$ behaves effectively as an inflation potential. This model provides novel insights into black hole interiors and, when extended to a cosmological setting, predicts a small but non-zero closed spatial curvature: $ -0.07 \pm 0.02 \le \Omega_k < 0$. This lower bound follows from the requirement of $\chi_k \ge \chi_* \simeq 15.9$ Gpc to address the cosmic microwave background low quadrupole anomaly.
The bounce remains confined within the initial gravitational radius $r_{\rm S} = 2GM$, which effectively acts as a cosmological constant $\Lambda$ inside $r_{\rm S}=\sqrt{3/\Lambda}$ while still appearing as a Schwarzschild black hole from an external perspective. This framework unifies the origin of inflation and dark energy, with its key observational signature being the presence of small but nonzero spatial curvature, a testable prediction for upcoming cosmological surveys.
\end{abstract}

\maketitle

\section{Introduction}
\label{S:1}

The standard model of cosmology, rooted in the Big Bang paradigm and the theory of General Relativity (GR) with the addition of cosmological constant ($\Lambda$) and cold dark matter (CDM), has been remarkably successful in explaining key observations, such as the Cosmic Microwave Background (CMB), the large-scale structure of the Universe, and the accelerating cosmic expansion attributed to dark energy. However, several fundamental questions remain unresolved, such as the nature of the initial singularity, the flatness and horizon problems (or the origin of inflation), and the physical origin of dark energy. These challenges have spurred the development of alternative frameworks that seek to extend or complement the standard cosmological model. The nature of CDM and the origin of the smallness of $\Lambda$ have been the central issues of modern cosmology. 

\begin{figure}
\centerline{
\includegraphics[width=50mm]{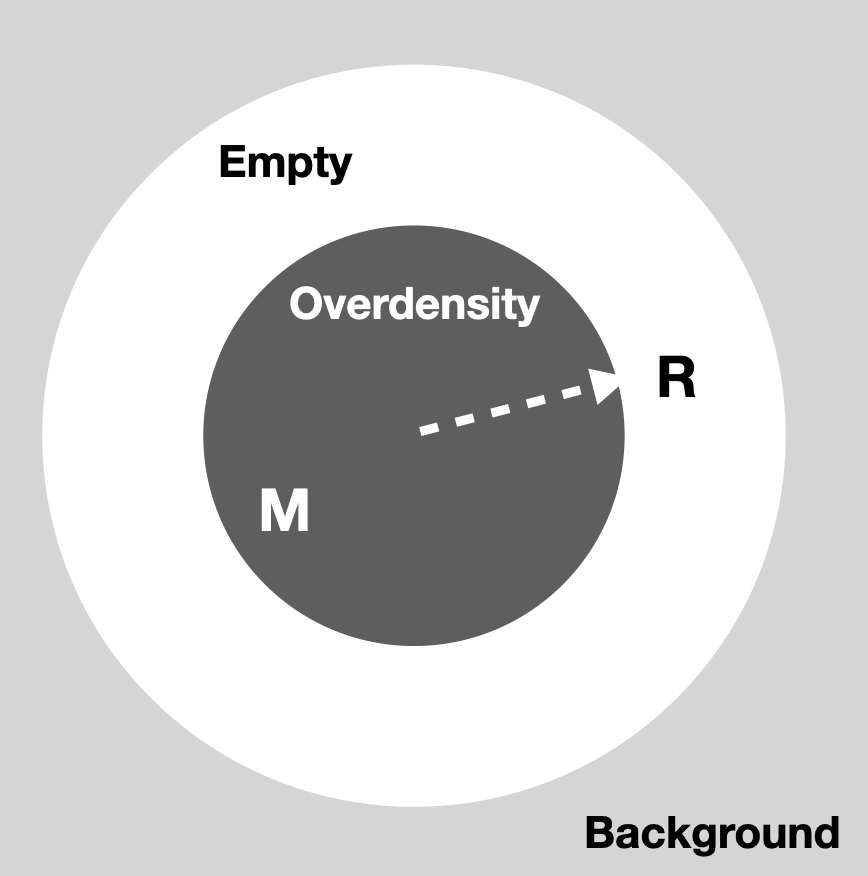}}
\caption{Graphical representation of the spherical collapse. There are three uniform spherically symmetric distributions: (i) outer background $\bar{\rho}$, (ii) inner region with radius less than $R$ and larger mean density $\rho>\bar{\rho}$, and (iii) empty space outside $R$. Two key ingredients of this configuration: a) the inner region collapse is decoupled from the background, and b) such gravitational bounded collapse is modeled in GR with a global spatial curvature $k>0$.} 
\label{fig:collapse}      
\end{figure}

A related problem is that of understanding the singular collapse into a Black Hole (BH). Both problems can be addressed when we consider the relativistic spherical collapse of a local (finite) Friedmann-Lemaitre-Robertson-Walker  (FLRW) cloud within a larger background, as shown in Fig.\ref{fig:collapse}. 
A cosmological bounce and an inflationary phase can emerge naturally from two fundamental assumptions: $ \dot{\rho} = 0 $ and $ k > 0 $, which simply follow from considering a finite over-density of matter obeying the quantum exclusion principle, which prevents the density from overcoming some threshold value or ground state $\rho_{\rm G}$.

Crucially, this quantum mechanism violates the strong energy condition (SEC) in classical GR. As was shown in \cite{Pradhan}, the bounce requires a non-zero local curvature $k>0$. Both conditions sidestep the singularity GR theorems proposed by \cite{Hawking_Penrose_Singularity}, allowing us to formulate a novel solution to a pivotal issue in cosmological theory. 

The bouncing scenario we formulate naturally extends to the subsequent stage of inflationary expansion when spatial curvature effects become negligible, resulting in the resolution of the horizon and flatness problems. Since the model is confined to a finite comoving region of spacetime, it introduces a finite comoving cutoff for super-horizon perturbations, potentially explaining anomalies in the CMB, such as the absence of structures beyond 66 degrees \citep{2022PhLB..83537468G}. This is a unique aspect of the model that is actually compatible with CMB observations, unlike the standard framework of inflationary cosmology.
The value of this cut-off directly relates to a prediction of spatial curvature $k>0$.

A recent paper (\cite{Pradhan}) has numerically solved the Newtonian spherical collapse equations with a polytropic equation of state (EoS) inspired by neutron star (NS) conditions. It found bounces at or above nuclear saturation density with equivalent GR behavior in a closed FLRW metric. The GR bounce corresponds to the ground state of the matter, characterized by \(P = -\rho\) (which is often termed a meta-stable state or quasi-deSitter in the framework of standard inflation). Here, we elaborate on the underlying mechanisms of this phenomenon and its implications for cosmic evolution, and we find new analytical and numerical solutions for the bounce, which are fully relativistic and within classical GR with a perfect fluid $P=P(\rho)$.

{
Our approach builds on the classical general relativistic treatment of a uniform, finite FLRW fluid ball (also referred to as a patch or cloud) embedded in a surrounding Schwarzschild vacuum spacetime, as studied by \cite{1963ZelDovich,2018arXiv180102249G,faraoni}. The dynamics of such configurations—in both expansion and collapse—have been examined in various contexts, including early treatments of relativistic collapse by Thompson and Whitrow \cite{1968ThompsonWhitrow}, and Bondi \cite{bondi}, who focused on pressure-supported configurations evolving toward black hole formation. Further developments by Smoller and Temple \cite{2003SmollerTemple} provide exact solutions featuring spherical shock waves, extending the Oppenheimer–Snyder model \cite{1939OppenheimerSnyder} to cases with non-zero pressure.}

{Some earlier literature suggests that matching a FLRW interior to a Schwarzschild exterior requires vanishing pressure—i.e., $P = 0$ at the junction—to ensure metric continuity across the density discontinuity. However, this restriction arises under the specific assumption that the spherical junction radius must follow a  geodesic of the FLRW metric, typically parametrized as $R = a(\tau) \chi_*$ with constant $\chi_*$. In the presence of pressure ($P \neq 0$), the radius of the junction is no longer a fixed comoving radial coordinate but evolves dynamically, i.e., $\chi_* = \chi_*(\tau)$, as shown explicitly in \cite{gaztanaga:bhu1}. In this case, a smooth matching of the FLRW and Schwarzschild metrics is still possible without introducing any discontinuity or requiring a surface layer with additional energy-momentum content.}

{
By contrast, a distinct class of solutions—often called ``bubble" or ``baby universe" models—assumes a vacuum de Sitter interior matched to a Schwarzschild exterior \cite{Gonzalez-Diaz,1987PhRvD..35.1747B,1989PhLA..138...89G,1989PhLB..216..272F,Aguirre,PBH3}. These spacetimes are inherently discontinuous: a smooth junction between two distinct vacuum solutions (de Sitter and Schwarzschild) is not possible without an intermediate surface layer carrying matter or tension, often modeled via a thin shell or bubble wall. Gravastar models \cite{gravastar2015} are examples of such constructions. Our setup, in contrast, considers a physically continuous metric across the boundary of a matter-filled collapsing region embedded in vacuum. No additional surface term or exotic matter layer is required, making it a straightforward realization of the classical spherical collapse framework.}

{Naturally, our proposed model is also related to existing cosmological bouncing scenarios (such as \cite{AASB,Starobinski1980,nojiri2017,2017PhR...692....1N,2020PhRvD.102j4042O,Kiefer:2024okk}).  Several classes of such bouncing cosmologies have been proposed in the literature. One prominent class involves modified gravity theories, such as $f(R)$ gravity and Gauss–Bonnet extensions, which allow for non-singular solutions that violate the strong energy condition (SEC) within a classical framework \cite{nojiri2017, faraoni2015}. These models produce bounces through geometrical modifications of Einstein’s equations, but often require fine-tuning and may suffer from instabilities or lack of a clear quantum limit. Another approach is rooted in applying loop quantum gravity techniques to homogeneous and isotropic spacetimes. In such a loop quantum cosmology, quantum corrections to the Friedmann equations introduce a repulsive force at Planckian densities, leading to a robust bounce that replaces the classical singularity \cite{bojowald2008}. This framework offers a well-defined and non-singular evolution at Planckian densities, but relies on the validity of the quantum gravity formalism. A third class involves effective matter models, in which exotic scalar fields produce violations of the SEC necessary for a bounce. These kinds of models employ scalar fields with negative kinetic terms or non-minimal couplings to achieve a non-singular evolution \cite{brandenberger2017, nojiri2017, cai2014} at the cost of introducing new degrees of freedom with ambiguous physical interpretation or require extensions beyond standard quantum field theory.
More recent developments include bounces arising from non-local gravity \cite{Biswas_2006, Sravan2}, string-inspired cosmologies \cite{gasperini2003}, and non-singular anisotropic models \cite{barrow2004, Sravan1}. Each of these introduces distinct mechanisms for evading singularities. However, new complex or speculative ingredients are required in these models.
Despite the diversity of all these approaches, a common limitation is the reliance on mechanisms beyond classical GR and standard matter physics. In this context, our Black Hole Universe (BHU) model, presents a novel framework where the observable Universe emerges from the gravitational collapse of a finite, nearly homogeneous FLRW cloud of ordinary matter in standard GR.}
It is worth noting here that when we include quantum effects in curved spacetime, even if we initially start with classical fluid with positive pressure, one gets negative pressure contributions as is shown in the studies of semi-classical gravity \cite{CesareSilva:2024nym,CesareSilva:2025kau,Shapiro:2008sf}.
The quantum exclusion principle sets a new universal goal for the theories of quantum gravity \cite{Loll:2022ibq,Buoninfante:2024yth,Buoninfante:2022ykf}. 
Although the full study of quantum gravity is beyond the scope of this current study, it lays a foundation for what the actual theory of quantum gravity is supposed to achieve or to be consistent with. However, a pre-requirement for a complete theory of quantum gravity is the robust development of quantum field theory in curved spacetime \cite{Kumar:2023ctp,Kumar:2023hbj,2836794}, which we aim to develop in the future for the non-singular origin of the Universe we present here.

The FLRW cloud bounce and subsequent inflation are driven by the degenerate pressure  $P =-\rho$, and this is supported by the hypothesis of the quantum exclusion principle, i.e., GR with quantum matter avoids singularities, which aligns very well with Misner's thoughts on how quantum theory should avoid singularities \cite{Misner:1969qxx}. 
A new key ingredient for our approach is to consider a finite cloud, which allows us to
incorporate spatial curvature, demonstrating its essential role in enabling the bounce. Finally, we address how cosmic acceleration emerges as a natural consequence of the bounce.

This work stands on its own, but it can also be used to extend and complement the Black Hole Universe (BHU) model (\cite{gaztanaga:bhu,gaztanaga_mou}) with a closed FLRW cloud $k>0$. The flat case $k=0$ used previously is a good approximation all the way to the point where we approach the singularity but does not allow for a bounce to occur. 
By connecting the early and late phases of cosmic evolution, it provides a unified model that bridges gravitational collapse, cosmic inflation, and the present accelerated expansion of the Universe. Grounded in physical principles and supported by numerical simulations, this model offers a compelling alternative to the standard cosmological paradigm while addressing its unresolved challenges.

{In this paper, we elaborate on the unified scenario in which the observable Universe originates from the gravitational collapse and bounce of a finite, closed FLRW cloud of matter. We begin by setting up the relativistic spherical collapse model, emphasizing the critical role of positive spatial curvature ($k>0$) in describing a bounded perturbation within a larger background. In the following section, we generalize our scenario to non-vanishing pressure and introduce a pressure contribution motivated by the quantum exclusion principle. This additional (degeneracy) pressure is related to a ground state with a constant energy density and ultimately prevents the singular collapse.}

{Building on this, the next sections we describe how degeneracy pressure halts the collapse, leading to a gravitational bounce without requiring modifications to GR and what the implications for cosmic inflation and acceleration are. In detail, we derive the exact analytical solution describing the bounce, demonstrate how the bounce smoothly evolves into an inflationary phase, providing a natural origin for cosmic inflation within the same relativistic framework and extend the model to late-time cosmology. There we discuss how the finite mass and size of the FLRW cloud imply an effective cosmological constant, offering a physical interpretation for $\Lambda$ in terms of the BHU model. Through this sequence, we show how gravitational collapse, bounce, inflation, and dark energy can be understood as different phases of a single, continuous process, rooted in classical GR combined with quantum mechanical principles.}
We use $c=1$ except when otherwise stated.

\section{Spherical collapse $P=0$}
\label{sec:P=0}

Here, we want to model the collapse of a finite cloud or perturbation within a larger background. We will assume that the initial cloud is a spherical overdense region of a perfect fluid that is surrounded by an empty space, as shown in Fig.\ref{fig:collapse}. This configuration is embedded in a larger volume containing a homogeneous, more diluted fluid. We also assume that $\Lambda=0$ or negligible to start with. For an observer moving with a perfect fluid, the energy-momentum tensor is diagonal: $\boldsymbol{T}_\mu^{~\nu} = diag[-\rho, P, P, P]$, where $\rho=\rho(\tau,\chi)$ is the relativistic energy density and $P=P(\tau,\chi)$ is the pressure.
The cloud is initially very large and has a very low density, so the pressure and temperature can be neglected. The relativistic solution to this problem was given by \cite{Lemaitre:1933gd} `atom universe' and is known today as the Lemaitre-Tolman-Bondi (LTB) model. The most general spherically symmetric metric in  the comoving frame (i.e., moving with the fluid) is:
{
\beq
ds^2 = \boldsymbol{g}_{\mu\nu} d\boldsymbol{x}^\mu d\boldsymbol{x}^\nu =- d\tau^2 + \left(\frac{\partial r}{\partial \chi}\right)^2 \frac{d\chi^2}{F^2(\chi)} + r^2 d\Omega^2,
\label{eq:LTB0}
\eeq 
where $F(\chi)$ is an arbitrary function of $\chi$. For the flat geometry, we have $F^2=1$. For the closed geometry case 
we have $F^2=1-k\chi^2$:
\beq
ds^2 =- d\tau^2 + \left(\frac{\partial r}{\partial \chi}\right)^2 \frac{d\chi^2}{1-k\chi^2} + r^2 d\Omega^2,
\label{eq:LTB}
\eeq 
with $k<1/\chi^2$.}
The physical radius (or aerial coordinate) $r=r(\tau,\chi)$ corresponds to the area distance and reflects spherical symmetry. The functional form of $\boldsymbol{g}_{\chi\chi}$ results from $\boldsymbol{T}_0^1=\boldsymbol{G}_0^1=0$ in the comoving frame.
The general solution to the Einstein field equation is:
\bea
H^2 &\equiv& \left(\frac{\dot{r}}{r} \right)^2 = \frac{2GM(\tau,\chi)}{r^3} -\frac{k\chi^2}{r^2}  \,, 
\label{eq:Hubble-LTB}
\\
M(\tau,\chi) &\equiv& 4\pi \int_0^\chi \rho(\tau,\chi^\prime) \, r^2 \, \frac{\partial r}{\partial \chi^\prime} \, d\chi ^\prime
\label{eq:active-mass}
\\
\dot{\rho} &=& - 3 H \rho\,,
\eea
where the over dots correspond to partial time derivatives.
The mass-energy $M$ was introduced by \cite{Lemaitre:1933gd} and
is sometimes called the active gravitational mass and coincides with the relativistic Misner-Sharp mass (\cite{Misner-Sharp}), which is defined for the more general case with pressure \cite{2015GReGr..47...84F}.

Assuming a homogeneous cloud $\rho=\rho(\tau)$ requires $H$ to also be homogeneous. Consequently, $r= a(\tau) \chi$ and $k$ has to be constant. We then have:
\bea
H^2 =  \left(\frac{\dot{a}}{a} \right)^2 = \frac{8\pi G}{3} \rho -\frac{k}{a^2}  
\label{eq:Hubble-LTB2}  &;& \rho = \rho_i \, \left(\frac{a}{a_i}\right)^{-3}\,,
\\ 
M(\chi) =  \frac{4\pi}{3} r^3  \rho  \quad ; \quad   m \equiv M(\chi_*) & = & 
\text{constant}\,,
\label{eq:active-mass2}
\eea
where $\rho_i=\rho(a_i)$ and $\chi_*=\chi(a_i)$ are the initial density and comoving radius of the cloud so that the mass $m$ inside $R=a\chi_*$ remains constant. Note that we use comoving units such that $a=1$ at present. This is the same solution as the FLRW solution, as expected. In general, we can choose $k$ to have any sign depending on the initial conditions. The case of interest here
is $k \equiv 1/\chi_k^2$ with $\chi_k>\chi_*$, which corresponds to an overdensity. The value of $\chi_k$ relates to the initial velocity $H_i \equiv H(a_i)$ of the cloud when $R=R_i \equiv a_i \chi_*$:
\beq
 1/(a_i\chi_k)^2 = 2Gm/R_i^3 - H_i^2
 \label{eq:chik}
\eeq
{This reproduces the well-known result that a closed FLRW model exactly mirrors the relativistic spherical collapse model (see §87 in \cite{Peebles1980}).
The empty region around the collapsing overdense perturbation separates the perturbation from the background and will expand with time. The corollary to Birkhoff’s theorem (the relativistic version of Gauss law) ensures that the spherical collapse evolution does not couple to the exterior spherically symmetric background (see \cite{faraoni}).}
In the Newtonian approximation, positive curvature ($k > 0$) corresponds to a system with negative total energy, where gravitational attraction exceeds kinetic energy, leading to the collapse of the cloud under its own gravity. In GR, spatial curvature provides the geometric representation of a gravitationally bound system. Just as a bound orbit in Newtonian gravity (like a planet around a star) is confined, a closed universe (or a collapsing region) in GR is “confined” by its own curvature. The metric of our initial perturbation
for $\chi<\chi_k$
is therefore the same as the one of a closed FLRW metric:
\beq
ds^2= \boldsymbol{g}_{\mu\nu} d\boldsymbol{x}^\mu d\boldsymbol{x}^\nu = 
-d\tau^2 + a^2 \left(\frac{d\chi^2}{1-k\chi^2} + \chi^2 d\Omega^2\right) \,.
\label{eq:FLRWk}
\eeq
Note that because $\chi_k$ is cosmologically large,
the corresponding curvature term $k/a^2=1/(a^2 \chi^2_k)$ is subdominant until $a$ becomes sufficiently small $a \rightarrow 0$.
{
Note that the solutions in Eqs.\ref{eq:Hubble-LTB2}-\ref{eq:active-mass2}, and also for Eqs.\ref{eq:Hubble-LTB}-\ref{eq:active-mass}, are the same as those in the Newtonian spherical collapse studied in \cite{Pradhan}. The FLRW cloud is a local and finite LTB solution in contrast to the standard FLRW metric, which is usually assumed to be global and infinite.
For $\chi<\chi_*$, both solutions are the same because of the corollary to  Birkhoff’s theorem ( see \cite{faraoni}).}

What happens in the LTB solution for $\chi>\chi_*$ in the region of empty space $\rho=0$ surrounding $R$?
Lemaitre also found a solution to this question. In \S11 of \cite{Lemaitre:1933gd}, he shows how variables can be changed to transform the LTB metric
into the static Schwarzschild metric. This change of variables corresponds to a rest frame, that is not comoving with the fluid, just as in the case of the static version of the de-Sitter metric (i.e., see \cite{decceleration}).
Another way to approach this question is to show that the FLRW metric matches the Schwarzschild metric without discontinuities {in agreement with the junction conditions}. 
Two different versions of this
approach were presented in \cite{BH_interior_Stuckey} and in \S 12.5.1 in \cite{padmanabhan_2010}. Such matching solution is what we call the FLRW cloud, which has the FLRW metric inside $\chi<\chi_*$ and the Schwarzschild metric outside $\chi_*$ (\cite{gaztanaga:bhu1} also presented the case $k=0$ and $\Lambda=0$ for timelike and null junctions).

{The relativistic spherical collapse can be interpreted  either as a solution to an LTB metric, as in Lemaitre (1933) original solution, or as a matching of two different solutions (as in the above references, see also Appendix A3 in \cite{2022PhLB..83537468G}). Both solutions are identical. In the LTB solution (i.e. Eq.\ref{eq:Hubble-LTB}-\ref{eq:active-mass2}) the density has a discontinuity at $\chi=\chi_*$:
\beq \rho(\tau,\chi) = \left\{ \begin{array}{ll}  \rho(\tau)  &  {\text{for}} ~ ~ \chi < \chi_*\\ 0  & {\text{for}} ~ ~ \chi \ge \chi_* \\ \end{array} \right. \,, \label{eq:rho1} 
\eeq
while keeping the spacetime metric continuous. The discontinuous density does not pose any problems here, because the extrinsic curvature still remains smooth. As was shown in \cite{gaztanaga:bhu1}, the two metrics can be matched without discontinuities. This is in contrast to, e.g., the concept of Gravastars, where the matching of a \textbf{(static)} de Sitter space inside with a Schwarzschild metric outside requires a particular treatment of the junction. The discontinuous $T_{\mu\nu}$ leads to a discontinuous extrinsic curvature, which needs to be cured by the introduction of an artificial thin shell of matter. This is not necessary in the BHU model.}

{Let us consider the fate of an FLRW cloud once the collapse has started. Analogously to any spherically symmetric matter distribution undergoing gravitational collapse, the outer physical radius $R=a(\tau) \chi_*$ of the cloud shrinks}
and eventually crosses inside the corresponding Schwarzschild radius: $R<r_{\rm S}=2G m$. When this happens, the FLRW cloud becomes a BH {\textbf{according to the exterior or superior observer}}. 
{ In contrast to standard collapse scenarios, where the mass involved is of stellar dimensions, here we consider cosmological scales. Consequently,}
both $r_{\rm S}$ and $a_i\chi_* \gg r_{\rm S}$ are {extremely} 
large, 
and the corresponding densities are {extremely} 
small.
{We can estimate} the density at a given time $\tau$ before the collapse to the singularity in the absence of any pressure 
\beq
\rho = \frac{\tau^{-2}}{6\pi G}
\simeq 3.97 \times 10^{-13}\, \frac{\rm M_{\odot}}{\text{km}^3} \,  \left[\frac{\tau}{\text{s}}\right]^{-2}\,.
\label{eq:rho}
\eeq
This is the solution to Eq.\ref{eq:Hubble-LTB2} when the spatial curvature term is neglected.
If we take $m$ to be as large as the mass of our observable Universe ($m \simeq 5\times 10^{22}\,\rm M_\odot$), we find from Eq.\ref{eq:active-mass2} that at horizon crossing, $r=r_{\rm S}= 2Gm/c^2$, the density of the BH, $\rho$ is
\beq
\rho_{\rm BH} = \frac{m}{\frac{4\pi}{3} r_{\rm S}^3} =\frac{3r_{\rm S}^{-2} c^2}{8\pi G}
\simeq 3.59 \times 10^{-48} \, \frac{\rm M_{\odot}}{\text{km}^{3}} \simeq 4 \, \frac{\text{protons}}{\rm m^3} \,.
\label{eq:BHrho}
\eeq
At these low densities, it is reasonable to assume that the thermal pressure $P$ and temperature $T$ are negligible because the time scale of the collapse is negligible compared to the time scales for any interactions between neutral particles. The cold collapse proceeds inside the BH event horizon. Note in Eq.\ref{eq:rho} how  even up to one second before the singularity occurs, the density is small compared to nuclear saturation density (SD) in atomic nuclei or  in NS: 
\beq
\rho(\tau=1s)\ll\rho_{\rm NS} \simeq \rho_{\rm SD} \simeq 1.4 \times 10^{-4}\, \frac{\rm M_{\odot}}{\text{km}^{3}} \,.
\label{eq:NS}
\eeq

\section{SPHERICAL COLLAPSE $P=P(\rho)$} 
\label{sec:P.neq.0}

{
So far, we have considered the spherical collapse of a FLRW cloud with a perfect uniform fluid and zero initial pressure, $P = 0$. As the collapse proceeds, interactions among particles lead to an effective pressure, modeled as $P = P(\rho)$. Due to the corollary of Birkhoff’s theorem (the relativistic version of Gauss law), 
the FLRW cloud solution inside $\chi_*$ will be the same as the one for the infinite FLRW  metric.}
The FLRW solution for the general case of $P= P(\rho)$, changes to:
\bea
H^2 =  \left(\frac{\dot{a}}{a} \right)^2 &=& \frac{8\pi G}{3} \rho -\frac{k}{a^2}\,,
\label{eq:Hubble-LTB3}
\\
\frac{\ddot{a}}{a} = - \frac{4\pi G}{3} (\rho + 3P)  &;& 
 \dot\rho + 3~ H~(\rho+P) = 0 \,,
\label{eq:ddota}
\eea
where we have set $\Lambda=0$ for clarity. Even if $\Lambda$ is non-zero, its contribution can be neglected when we approach high densities.

\section{Degeneracy Pressure} 
\label{sec:degeneracy}

Here, we draw an analogy of our understanding of the Universe with NSs and astrophysical black holes.
As the collapsing cloud approaches the singularity ($a \rightarrow 0$), the density $\rho = a^{-3} a_i^{3} \rho_i$ increases without bound. However, once any fermionic constituent of the cloud reaches its quantum ground state, the Pauli Exclusion Principle generates a degeneracy pressure, $P = P(\rho)$, independent of temperature. Remarkably, this degeneracy pressure and the corresponding equilibrium density apply universally to systems ranging from atoms to NS despite their vast difference in mass—approximately $10^{57}$ times.
For even larger masses, such as the mass of the Universe (about $10^{22}$ times greater than that of a NS), the degeneracy pressures of electrons, neutrons, or even quarks may not suffice to halt the collapse. Indeed, for masses exceeding the Tolman–Oppenheimer–Volkoff (TOV) limit of 2–3 $\rm M_\odot$, a black hole forms, and the collapse proceeds within the event horizon, leaving the internal physics largely unexplored. 
A version of the Pauli Exclusion Principle should remain valid even under extreme conditions, as no two fermions can occupy the same quantum state. Thus, a new quantum ground state, characterized by a maximum density $\rho_{\rm G}$, could also emerge if electrons and quarks are not fundamental, preventing a true singularity. This notion lies at the heart of applying principles of quantum theory in the context of gravity, which offers a framework to circumvent singular collapse and explore the limits of physical laws in extreme conditions. 



In the central regions of the collapsing cloud, where the bounce occurs, the pressure and density can be treated as approximately uniform in the comoving frame. The validity of this assumption was demonstrated in \cite{Pradhan} using hydrodynamical simulations.

We can see from Eq.\ref{eq:ddota} that as $\dot\rho  \rightarrow 0$, the relativist pressure $P \rightarrow -\rho$. Appendix A shows one way to understand this in terms of scalar fields, where the EoS plays the role of the scalar potential $V(\phi)$.

We can define a cloud radius $R_{\rm G}$  from Eq.\ref{eq:active-mass2}:
\beq
\frac{8\pi G \rho_{\rm G}}{3} \equiv  \frac{r_{\rm S}}{R_{\rm G}^3} \,.
\label{eq:RG}
\eeq
which corresponds to the radius $R_{\rm G}$ of the cloud when it reaches $\rho=\rho_{\rm G}$ if we neglect the effects of pressure.
For the mass of the Universe $m= 5\times 10^{22}\, \rm M_\odot$ and when assuming nuclear saturation density (SD) as a lower limit $\rho_{\rm G}>\rho_{\rm SD}$ (Eq.\ref{eq:NS}):
\beq
R_{\rm G}< r_{\rm SD} =  4.4 \times 10^{8} \, \text{km} \simeq 1.43 \times 10^{-14} \, \rm Gpc\,.
\label{eq:NS2}
\eeq
This value of $R_{\rm G}$ represents the beginning of the transition into the ground state. The model transitions from a state of constant total energy-mass (with a uniform but evolving energy density) to a state of uniform and time-invariant energy density.

We have found that the quantum exclusion principle leads to a ground state where the relativistic equation of state (EoS) becomes: \( P = -\rho \). This EoS is fundamentally distinct from the one typically considered under nuclear saturation in NSs. A key assumption for NS is that GR is negligible at scales of inter-quantum interactions.
However, it is crucial to recognize that gravity is inherently nonlinear. The active mass-energy, as defined in Eq.\ref{eq:active-mass}, includes not only matter but also gravitational energy, which can not be neglected.
 This behavior is also captured in the relativistic continuity equation (Eq.\ref{eq:ddota}): 
\beq
\dot\rho =- 3H(\rho + P/c^2),
\label{eq:rhodot}
\eeq
where we have included $c$ to illustrate that the second term  $P/c^2$ is purely relativistic and does not appear in the Newtonian equation. This equation demonstrates that once a constant density is reached, the EoS naturally transitions to \( P = -\rho \) (back in units of $c=1$). This behavior is not captured by Newtonian dynamics (or relativistic corrections) and is, therefore, not present in conventional EoS models for NS, emphasizing the necessity of considering relativistic effects in describing such ground states. In the Newtonian approach, the pressure only appears as a force in the Euler equation. The GR analog of the Euler equation (or its first integral) is the Hubble-Lemaitre law in Eq.\ref{eq:Hubble-LTB3}, which is independent of pressure. The two approaches come together when we combine the Euler equation with the continuity equation.

The other important difference between our comoving EoS and the NS modeling is that we are considering the collapse (and later expanding) phases and not a static solution. So, our EoS lives in the comoving frame, while NS EoS refers to a Newtonian rest frame. What does the relativistic EoS look like in the rest frame? This is presented in Appendix B.

\section{Bouncing solution}

Bringing together the insights from the two previous sections, we can now explore what happens during the collapse of an FLRW cloud as it reaches its ground state density somewhere above nuclear saturation.
For a constant $P=-\rho = -V_{\rm G}$,
Eq.\ref{eq:Hubble-LTB3}-\ref{eq:ddota} become:
\bea
\frac{\ddot{a}}{a} &=& + \frac{8\pi G}{3} \rho_{\rm G} \equiv \frac{r_{\rm S}}{R_{\rm G}^3} \quad ; \quad 
  \dot{\rho}_{\rm G} = 0 \,,
\label{eq:ddota2}
\\
H^2 &=&   \left(\frac{\dot{a}}{a} \right)^2 =\frac{r_{\rm S}}{R_{\rm G}^3} -\frac{k}{a^2}\,.
\label{eq:Hubble-LTB4}
\eea
This corresponds to a gravitational bounce ($\dot{a}=0$ and $\ddot{a}>0$) at:
\beq
a_{\rm B} = \sqrt{\frac{R_{\rm G}^3}{\chi_k^ 2 r_{\rm S}}}   \quad \text{or} \quad
R_{\rm B}^2 = R_{\rm G}^3/r_{\rm S}
\,.
\label{eq:a_B}
\eeq
Note how it is critical that $k>0$ (or $\chi_k^2<\infty$) to have a bounce before the singularity ($a=0$) occurs. The bounce is only possible because
both $R_{\rm G}>0$ and the cloud is finite (that is, $\chi_k<\infty$ and $r_{\rm S}<\infty$). It is physically inconsistent to perceive a bouncing scenario in an infinite FLRW cloud.  
This is reflected above by the mathematical fact that an infinite FLRW cloud has no bounce. 

The exact solution to Eq.\ref{eq:Hubble-LTB4} is:
\beq
a = \frac{a_{\rm B}}{2}  \left[ e^{-|\Delta \tau|/R_{\rm B}}+e^{+|\Delta \tau|/R_{\rm B}} \right]
= a_{\rm B} \cosh{(\Delta \tau/R_{\rm B})}\,,
\label{eq:bounce}
\eeq
where $|\Delta \tau|$ is the time to/from the bounce ($a=a_{\rm B}$) with $a>a_{\rm B}$, and $R_{\rm B}$ is the radius of the cloud when it bounces:
\beq
R_{\rm B} \equiv a_{\rm B} \chi_* = \sqrt{R_{\rm G}^3/r_{\rm S}} = \sqrt{\frac{3}{8 \pi G\rho_{\rm G}}} \,,
\eeq
which happens to be the gravitational radius of the ground state $\rho_{\rm G}$. 

\section{Gauss curvature scale}

Recall from Eq.\ref{eq:chik} that $\chi_k$ needs to be larger than the cloud boundary: $\chi_k>\chi_*$.
The existence of $\chi_*$ imposes a natural cutoff in the spectrum of super-horizon perturbations generated during collapse, bounce, or inflation. This cutoff shows up
in the CMB sky as:
\beq
\theta_{\rm cut} = \frac{\chi_*}{\chi_{\rm CMB}}\,,
\eeq
where $\chi_{\rm CMB} \simeq 13.8$ Gpc is the comoving radial distance to the CMB for $\Omega_\Lambda \simeq 0.7$ and $H_0 \simeq 70$ km/s/Mpc.
Strong evidence for such a cutoff has been known
since COBE and confirmed by WMAP and Planck (\cite{COBEw2,Bennett-wmap,P18isotropy}). Reference
\cite{CMB_cutoffs_Camacho} estimated the homogeneity scale to be
\beq
\theta_{\rm cut} \simeq 65.9 \pm 9.2 \, \text{degrees},
\label{eq:theta_cut}
\eeq
which implies:
\beq
\chi_* \simeq 15.93 \pm 2.22 \, \text{Gpc} \,,
\label{eq:chi_cut}
\eeq
which is the Gaussian curvature scale. This interpretation predicts a value $\Omega_k$ today ($a=1$) of:
\beq
\Omega_k  \equiv  -k\left(\frac{1}{H_0} \right)^2 
= -(0.07 \pm 0.02) \left(\frac{\chi_*}{\chi_k}\right)^2\,,
\label{eq:omegak}
\eeq
which, for $\chi_k>\chi_*$,  is consistent with a critical reanalysis of the Planck Legacy 2018 data \cite{2020NatAs...4..196D}. 
This result also agrees with a previous independent way of modeling the low quadrupole $C_2$ measured in the WMAP power spectrum \cite{2003MNRAS.343L..95E}. The limits for $\Omega_k$ above assume that the homogeneity scale is the result of only $\chi_*$. This also explains the low quadrupole $C_2$  \cite{CMB_cutoffs_Camacho}. However, if the homogeneity scale or the low value of $C_2$ has a different origin, then the value of $\Omega_k$ in the floating FLRW cloud could be smaller. Inflation preceded by a bounce requires $\Omega_k<0$, and this could be found in upcoming cosmic surveys, as indicated by the analysis in \cite{2020NatAs...4..196D}.

\section{Cosmic Inflation}

The solution in Eq.\ref{eq:bounce} corresponds to an exponential expansion (or collapse) after (or before) the bounce, leading to a de-Sitter phase, just as in standard cosmic inflation. As mentioned before, the EoS plays the role of the inflation potential, and the actual solution is quasi-de-Sitter as we approach the respective ground state of the matter.

As an example, we can take the following toy ansatz to interpolate from $\rho \simeq \rho_{\rm G} (a/a_{\rm G})^{-3}$ to $\rho \simeq \rho_{\rm G}$:
\beq
\rho_* \equiv \frac{\rho}{\rho_{\rm G}} = \frac{1}{1+ [(a-a_{\rm B})/a_{\rm G}]^3}\,,
\eeq
where $a=a_{\rm G}+a_{\rm B}$ at the time when the density is half of $\rho_{\rm G}$ and quantities with a star index ($_*$) are given in units of $\rho_{\rm G}$. Eq.\ref{eq:Hubble-LTB3}  becomes:
\beq
(\dot{a} \chi_k)^2 =  \frac{(a/a_{\rm B})^2}{1+ [(a-a_{\rm B})/a_{\rm G}]^3}  - 1 \,.
\label{eq:Hubble-mod-anstz}
\eeq
Solving this numerically, we can obtain the exact equation of state $P(\rho)$, using Eq.\ref{eq:ddota}. For single field inflation like Starobinsky: the number of e-folds is $N_e=2/(1-n_s) \simeq 57$ (where $n_s$ is the scalar spectral index) \cite{PlanckInflation}.  The current bound on the tensor-to-scalar ratio is $<0.028$  (\cite{Galloni:2022mok}). This solution corresponds to $a_{\rm G} \simeq 5.69 \times 10^{24} a_{\rm B}$ and is shown in Fig.\ref{fig:bounce}.
The corresponding EoS follows: 
\beq
P/\rho_{\rm G}=-(\rho/\rho_{\rm G})^2 \quad \text{or} \quad P_* = -\rho_*^2
\label{eq:EoSnum}
\eeq
which corresponds to the generalized Chaplygin gas with $\alpha=2$ \cite{2004PhRvD..69b3004M}.

The ground state is approached asymptotically, with the density remaining constant even as the scale factor grows or decreases exponentially. This behavior arises because the active mass $m$ is no longer constant, a purely relativistic effect where the gravitational field itself contributes non-linearly to the source term. Consequently, the model transitions from a regime of constant energy-mass, characteristic of a Newtonian solution, to one of constant energy-density, which is inherently relativistic.

As mentioned before, the saturation densities in NSs and the nucleus of an atom are comparable with Eq.\ref{eq:NS} despite the former having a mass $ 10^{57}$ times larger than the latter. However, as discussed in \cite{Pradhan}, the densities at which the condition $P=-\rho$ is fulfilled for masses much larger than that observed for NSs could be significantly higher than that of Eq.\ref{eq:NS} so that the energy $\rho_{\rm G}$ of the corresponding cosmic inflation could be much larger.  In Appendix A, we show a more detailed comparison of $P=P(\rho)$ with inflation parameters and CMB observations. The amplitude of CMB fluctuations relates to a ground state that has energy densities much larger than nuclear saturation. The quasi-scale invariant spectrum and quantum parity features observed in the CMB (see \cite{2024JCAP...06..001G}) will also be reproduced with our bouncing solution.

In summary, a bounce driven by degeneracy pressure could give rise to an epoch of cosmic inflation and reheating. This opens the possibility for an epoch of nucleosynthesis and recombination similar to that in the standard model (e.g., see \cite{Steigman}).
Note that in standard inflation, reheating requires an oscillatory scale factor around the matter-dominated phase after the exponential expansion; in typical single field inflation: $a(t)\sim t^{2/3}\LF 1+\frac{1}{Mt}\sin(Mt) \RF$ where $M$ is inflaton mass. The equivalent process in terms of $P=P(\rho)$ is detailed in Appendix A.
This process has the potential to enable nucleosynthesis and recombination in a manner similar to the standard Big Bang model. More importantly, it also provides an alternative framework for understanding both early and late-time cosmic acceleration.

\section{Cosmic Acceleration}
\label{sec:lambda}

There is compelling observational evidence that the cosmic expansion is accelerating: $\ddot{a} > 0$ \citep{Riess_1998, Schmidt_1998, Perlmutter_1999,2003ApJ...597L..89F, Eisenstein_2005, Betoule_2012}. This acceleration appears to be dominated by the cosmological constant $\Lambda$. The $\Lambda$ term can be interpreted either as a fundamental modification of General Relativity (GR), denoted as $\Lambda_{\rm F}$, or as an effective dark energy (DE) fluid, $\Lambda_{\rm DE}$, analogous to the ground state $\psi$ described earlier, but with a much smaller energy density, $\rho_{\rm G} = \rho_{\rm DE}$. 

Regardless of the interpretation, the corresponding characteristic length scale,
\beq
R_\Lambda = \sqrt{3/\Lambda},
\eeq
is vastly larger than the nuclear saturation scale $R_{\rm G}$ (i.e., $8\pi \rho_{\rm G} \gg \Lambda$). Consequently, $\Lambda$ can be neglected in our discussion of the gravitational bounce and the corresponding inflationary period.

The measured value of $\Lambda$ is extremely small but non-zero, and its fundamental origin remains an open question. Although its connection to the fundamental laws of physics is unclear, its effect is well understood: it induces an event horizon, $R_{\rm H}$, in FLRW space-time:
\beq
R_H = a \int \frac{d\tau}{a} = a\int \frac{da}{H a^2} < R_\Lambda,
\eeq
beyond which regions ($R > R_{\rm H}$) are not causally connected to the interior ($R < R_{\rm H}$).
The standard assumption in cosmology is that the Universe beyond $R_{\rm H}$ is identical to the interior. However, this assumption presents two fundamental issues:

\begin{itemize}
    \item \textit{Lack of causal explanation}: The standard approach cannot provide a mechanism to explain how the Universe could be the same beyond $R_{\rm H}$. Cosmic inflation does not solve this puzzle because even under exponential expansion $a \sim e^{\tau {\rm H}}$, we have that $R$ is always $R<R_{\rm H}$. This is because the comoving distance traveled by light during $\Lambda$ domination $\chi = \int_a^\infty \frac{da}{H a^2} = \frac{1}{aH}$ exactly cancels the exponential expansion in $R=a\chi$.
    \item \textit{Violation of the variational principle}: Einstein’s field equations require that the metric asymptotically approaches Minkowski space at large distances, which is not satisfied if the FLRW universe extends indefinitely.
{At any given cosmic time (e.g., the present), the FLRW universe has the same finite, non-zero uniform density everywhere, including at spatial infinity. Hence, it is not asymptotically Minkowski. Even when the matter-energy content vanishes with time, the asymptotic metric is de Sitter and not Minkowski.}
    
\end{itemize}

Instead, if we assume that the region beyond $R_{\rm H}$ is empty, both of these issues are resolved. This leads to a finite universe with size $R_{\rm H} \rightarrow R_\Lambda$ and a finite total mass $m$ contained within it. The exterior is then naturally described by the Schwarzschild metric. This agrees well with the FLRW cloud model presented here in previous sections and in Fig.\ref{fig:collapse}. For consistency, we need to identify $R_\Lambda$ with the Schwarzschild radius:
\beq
R_\Lambda = r_{\rm S} = 2Gm.
\eeq
as both quantities are constant.
This immediately provides a physical interpretation of $\Lambda$:
\beq
\Lambda = \frac{3}{r_{\rm S}^2}.
\eeq
Thus, $\Lambda$ simply corresponds to the total mass $m$ of our finite Universe. This also explains why $\Lambda$ is small but nonzero: it is directly linked to the total mass of the Universe as in our FLRW cloud model. Thus, the measurement of a  $\Lambda$ can be interpreted as a measurement of $m$ and a confirmation that we live within a large but finite FLRW cloud model. This is also consistent with our new interpretation of the origin of the bounce and cosmic inflation presented here. 

This reasoning provides a straightforward and intuitive explanation of $\Lambda$ without requiring detailed calculations. Such calculations are presented in \cite{gaztanaga:bhu1}. By applying the relevant matching conditions, it is found that the radial null geodesics $R_{\rm H}$ satisfy Israel’s matching conditions and that the action principle correctly includes the extrinsic curvature boundary term, $K = 2/r_{\rm S}$.

This boundary interpretation of $\Lambda$ corresponds to the Black Hole Universe (BHU) model. For observational values $\Omega_\Lambda \simeq 0.70$ and $H_0 \simeq 70$ km/s/Mpc, we obtain:
\bea
r_{\rm S} &=& \frac{c}{H_0 \sqrt{\Omega_\Lambda}} \simeq 5.1 \pm 0.1 \, \text{Gpc}, \\ 
m &=& (5.4 \pm 0.1 ) \times 10^{22}\,\rm M_\odot \,,
\eea
with uncertainties from \cite{2024arXiv240403002D}.
Note that $r_s< \chi_*$ in Eq.\ref{eq:chi_cut}, which indicates that the perturbation form before becoming a black hole.

\section{Discussion and Conclusion}\label{sec:conclusions}

\begin{figure*}
    \centering
        \includegraphics[width=0.32\textwidth]{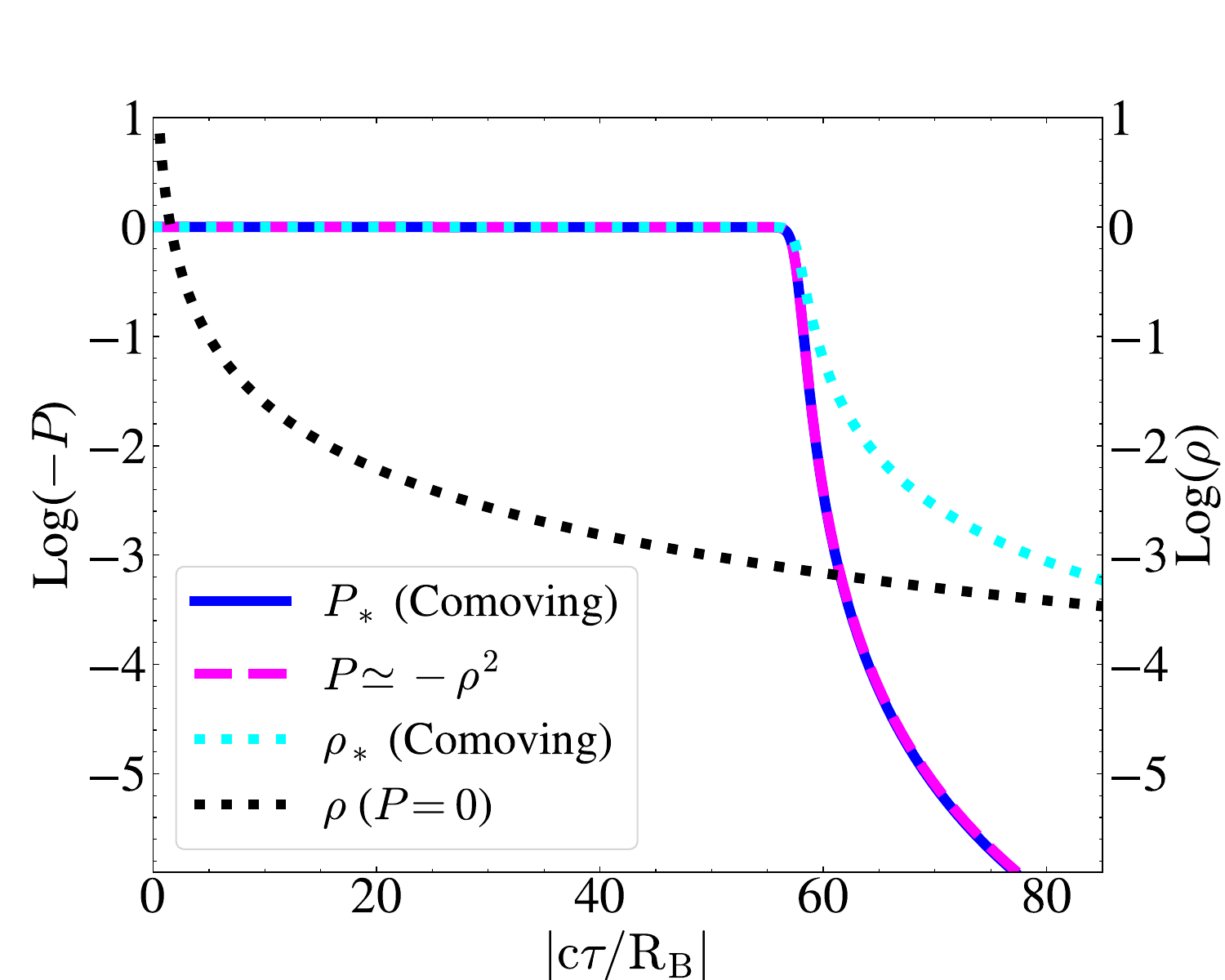}        
        \centering
        \includegraphics[width=0.32\textwidth]{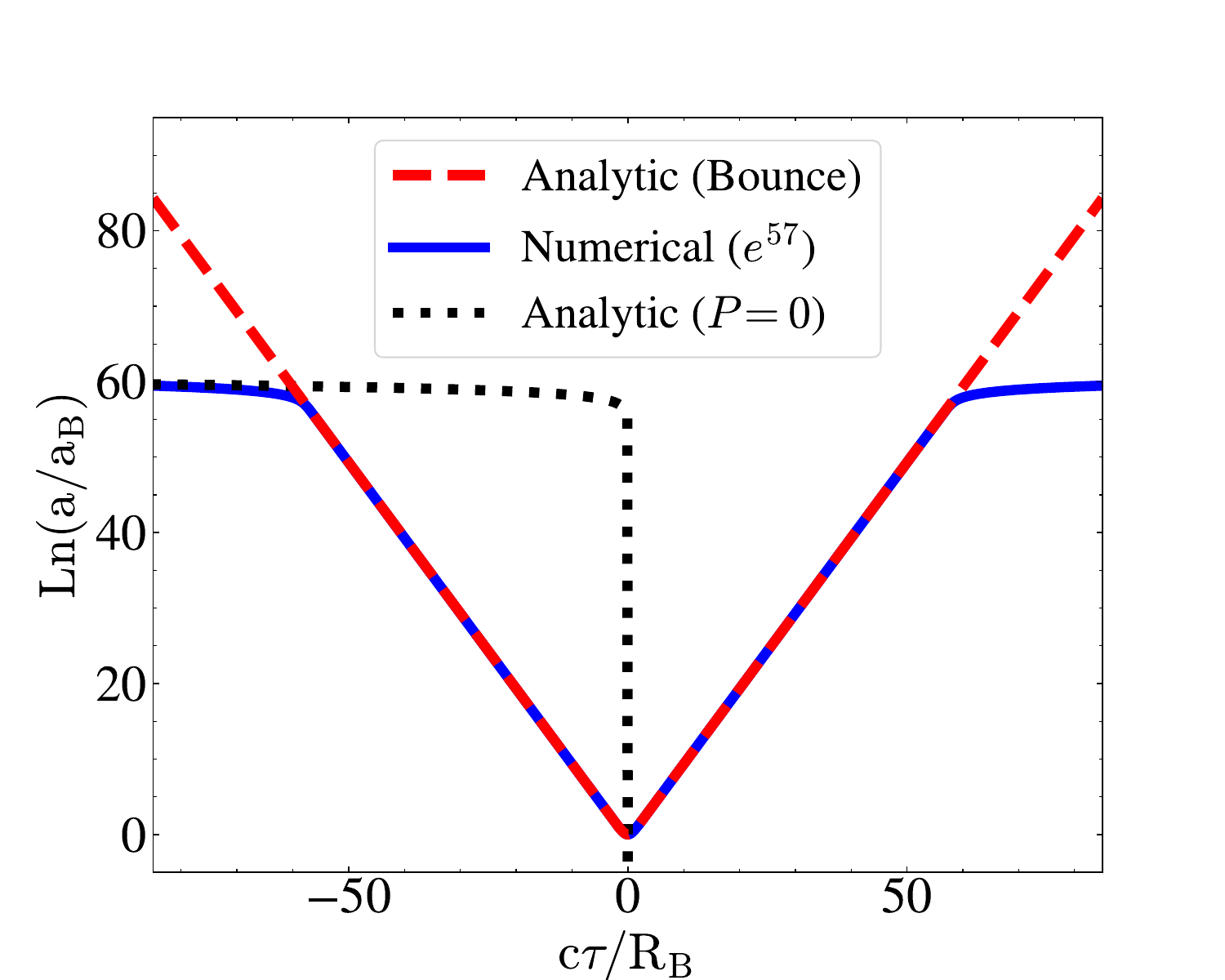}        
        \centering
        \includegraphics[width=0.32\textwidth]{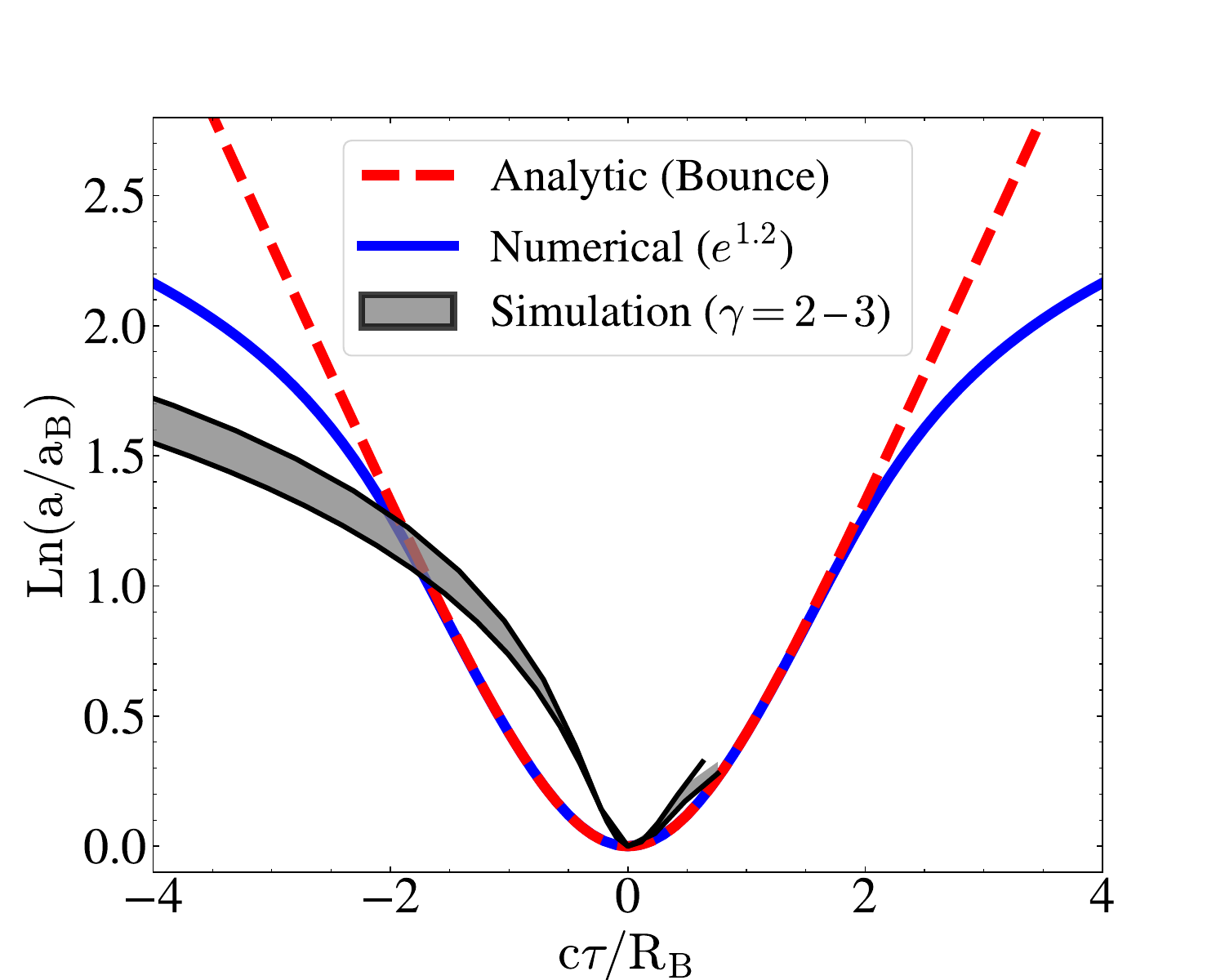}
    \caption{Left:  EoS ($P=P(\rho)$, blue, and $\rho$, cyan) in units of $\rho_G$,  derived using the numerical solution of Eq.\ref{eq:Hubble-mod-anstz} and Eq.\ref{eq:ddota}. These are compared with $\rho$ from the $P=0$ case (black dotted) and the polytropic fit $P_*=-\rho_*^2$ (dashed magenta). Middle: comparison of the different solutions for scale factor\textemdash the analytic solution presented in Eq.\ref{eq:bounce}, the numerical solution of Eq.\ref{eq:Hubble-mod-anstz} for 57 e-folds of inflation, and the singular pressureless solution denoted with dashed, solid and dotted lines respectively. Right: zoom-in around the bouncing region, where we compare the analytical bounce solution in Eq.\ref{eq:bounce} and the full numerical solution for $a(\tau)$ in Eq.\ref{eq:EoSnum} for 1.2 e-folds, with the Newtonian numerical simulations (\cite{Pradhan}), with different polytropic EoS for nuclear saturation in NSs: $P=K \rho^\gamma$ with $\gamma=2-3$.}
\label{fig:bounce}      
\end{figure*}

This paper presents a novel solution to the relativistic spherical collapse model for a bounded perturbation ($k > 0$). The key innovation lies in the introduction of a variable equation of state, $P = P(\rho)$, which asymptotically evolves from a pressureless, homogeneous state to a ground state characterized by a time-independent energy density. This transition naturally gives rise to a de Sitter phase in the final stages of collapse—immediately preceding the bounce—and persists throughout the ensuing expansion. The bounce itself admits an analytical expression, provided in Eq.~\ref{eq:bounce}.

The cosmological implication of this new approach is a novel understanding of the origin of the universe that emerges from the collapse and subsequent bounce of a spherically symmetric matter distribution. 
We show that upon reaching a quantum ground state, the relativistic matter equation of state (EoS) transitions from $P=0$ to $P = -\rho$ in the comoving frame. The relativistic degenerate pressure generated in this state halts the collapse and initiates a bounce and an inflationary expansion. We discussed how this mechanism parallels phenomena in NS physics and core-collapse supernovae (see reviews by \cite{bethe1990,burrows1995,Hebeler_2013,janka2012}), where the ground state is determined by nucleonic or quark interaction potentials within proto-NS. On the other hand, the expansion mechanism right after the bounce also parallels that of cosmic inflation, as detailed in Appendix A.

The non-singular bounce that happens inside a closed FLRW cloud (i.e., a finite-sized Universe trapped inside an event horizon) is induced by quantum matter EoS, which results in degenerate negative pressure.  
This same process drives an exponential expansion analogous to cosmic inflation, offering a novel solution to key challenges in standard cosmology, such as the origin of inflation and dark energy.
Our findings highlight the profound implications of relativistic quantum principles in shaping the early Universe. 

We start from a low-density cloud with  $a_i\chi_* \gg r_{\rm S}$ where $a_i \gg 1$ is the adimensional scale factor in units of the value today $a=1$. The key simplicity of this model is that the BH is not an ad-hoc initial condition to our system but a consequence of gravitational collapse. Without this, there is no argument for $\Lambda =3/r_{\rm S}^2$. This will happen for any initial condition where the initial density is sufficiently low.  Harrison-Zeldovich-Peebles \cite{Harrison1970,Zeldovich1970,Peebles1970} independently argued that gravitational instability alone (without inflation) would naturally produce a scale-invariant spectrum of perturbations out of an FLRW metric. Such perturbations will give rise to overdensities such as the ones considered here as the starting point to our Universe. This is our new guess for the ``initial condition".

\begin{figure*}
\centerline{\includegraphics[width=130mm]{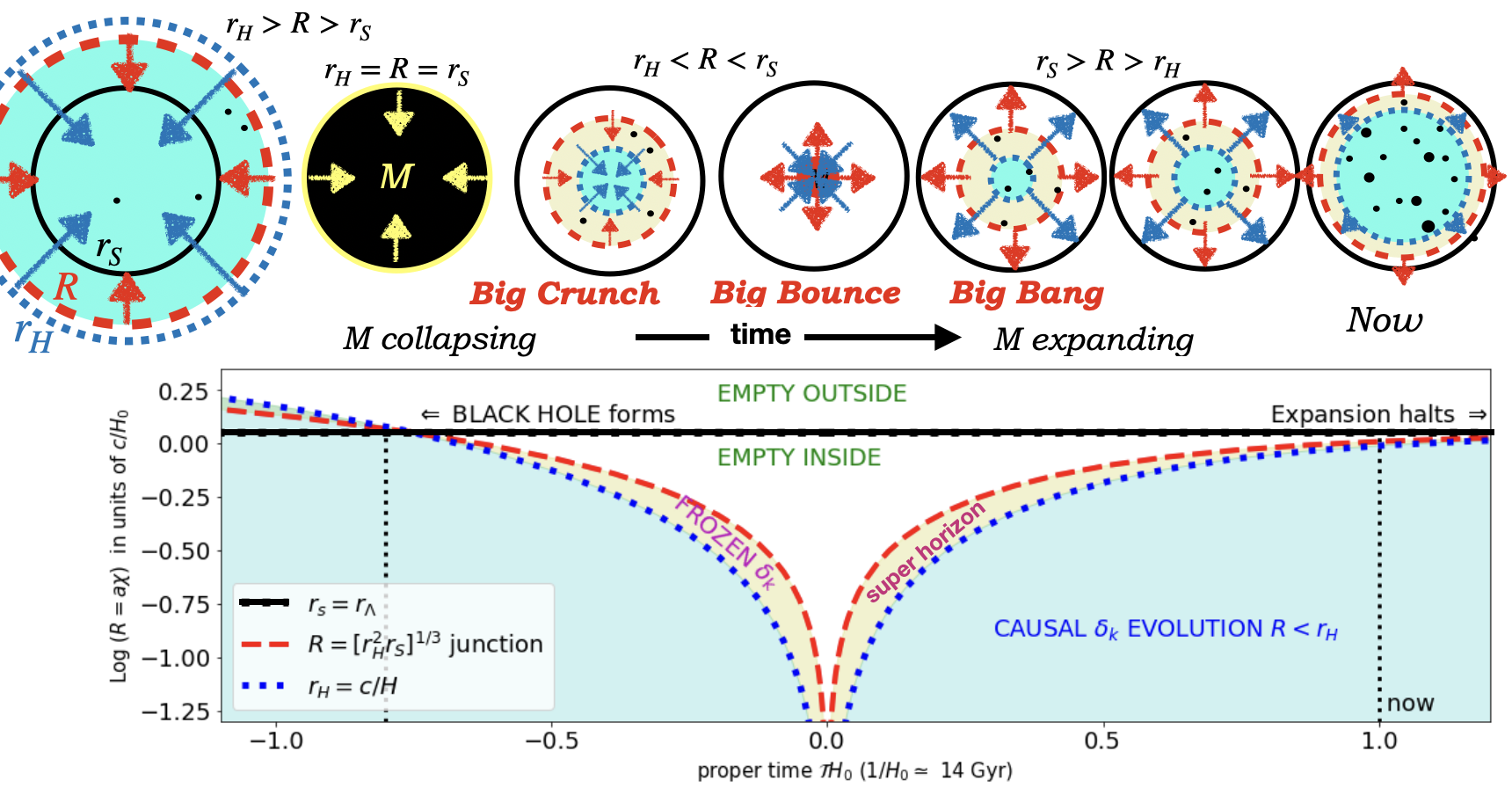}}
\caption{Time evolution in the radius of the FLRW cloud $R(\tau)= a(\tau) \chi_*$, first forming a BH and then bouncing inside to form our current observed expanding Universe. The black circle represents the FLRW cloud gravitational radius, which is also the asymptotic $\Lambda$ event horizon of our current expansion. The red dashed and blue dotted lines and circles correspond to the FLRW cloud radius and the Hubble radius $r_{\rm H}=c/H$. This figure has been adapted from \cite{gaztanaga:bhu}. Licensed under CC-BY.}
\label{fig:bounce2}      
\end{figure*}

Figure \ref{fig:bounce2} shows the full evolution of the finite FLRW cloud radius, \( R(\tau) \), illustrating how the horizon problem is resolved within the bounce model. 
In the middle panel of Figure \ref{fig:bounce}, we now present the exact bouncing solution, comparing the analytical expression for the scale factor \( a(\tau) \) given in Eq.~\ref{eq:bounce} with the numerical solution of Eq.~\ref{eq:Hubble-mod-anstz}. For reference, we also plot the singular analytic solution for \( P=0 \). Such singular behavior is unavoidable for an infinite cloud (\(\chi_* = \infty\) or \( r_{\rm S} = \infty \)) or in a flat or open geometry (\( k \leq 0 \)). Note how the inflationary phase has the right number of e-folds
$N_e =2/(1-n_s) \simeq 57$ consistent with the scalar spectral index $n_s$ measured by Planck \cite{PlanckInflation}.

It is evident that the analytic solution in Eq.~\ref{eq:bounce} is valid only for the inflationary phase, i.e., near the bounce, whereas the numerical solution accounts for the entire evolution, including the pre-bounce phase when \( P \approx 0 \). Before reaching the bounce, the numerical solution follows the pressureless analytic case closely, which is expected since the pressure is initially vanishing and then transitions smoothly into a constant (negative) degeneracy pressure (\( P_* = -\rho_* \)), as shown in the left panel. In this regime, the equation of state (EoS) is well approximated by \( P_* = -\rho_*^2 \) (dashed line). 
When we plot the pressure calculated for the numerical model (labeled $e^{57}$) as a function of the corresponding density and fit it, the fitting curve is $P=K \rho^\gamma$ with $K \simeq -1$ and $\gamma \simeq 2$.  This can be interpreted as a polytropic EoS with $\gamma=2$.

Furthermore, the transition point is clearly marked: as soon as pressure begins to build up, the numerical solution in the middle panel shifts from the analytic \( P = 0 \) solution (dotted line) to the analytic bounce solution (dashed line), denoting exponential collapse and vice-versa for exponential expansion. 

The right panel provides a zoomed-in view of the bounce region, where we compare our asymptotic analytical solution with the numerical Newtonian simulations of \cite{Pradhan}, which adopt an equation of state of the form \( P = K \rho^\gamma \). This type of EoS serves as a reasonable approximation for nuclear degenerate matter, with \( \gamma = 2 \) to \( 3 \), in the Newtonian framework \cite{Lattimer_Prakash_eos}. While the Newtonian simulations remain an approximation, we observe that both models yield a strikingly similar exponential expansion post-bounce, as also noted by \cite{Pradhan}. Note that the Newtonian simulation results presented here are for a 20 M$_\odot$ cloud, for which the bounce occurs at around nuclear saturation densities and the expansion has $\approx 1.2$ e-folds. For larger masses, we will have a larger number of e-folds as in the middle panel for the mass of our Universe.

The analytic solution we found in Eq.\ref{eq:bounce} is one of the cases considered in Eq.7 in \cite{Starobinski1980}, which corresponds to a de-Sitter Universe with closed curvature.
Instead of degeneracy pressure, this model arises from quadratic curvature modification of the Einstein-Hilbert action motivated by 1-loop self-energy contributions due to quantum matter, which leads to the first model of cosmic inflation (\cite{inflation_starobinsky}). 
But the reason to consider closed curvature (other than to produce a bounce as in \cite{AASB}) is not clear in this model. 
In the BHU model, the spatial curvature naturally results from the spherical collapse of a large overdensity confined to a finite region of spacetime.

Figure 3 in \cite{gaztanaga:bhu1} illustrates how the boundary $R(\tau)$ of the FLRW cloud is always outside the observational window for any (off-centered) observer inside the cloud. 
This is a general property of quasi-de-Sitter space and implies that the BHU does not result in observed anisotropies in the background of the cloud boundary. But the bounce and the initial cloud's comoving radius $\chi_*$
can result in a cutoff of the super-horizon quantum perturbations generated during inflation, which can be observed in the CMB \citep{CMB_cutoffs_Fosalba_2021,CMB_cutoffs_Camacho,2022PhLB..83537468G} and results in the constraint given by Eq.\ref{eq:chi_cut}. Such cutoff, together with parity asymmetry, predicts a lower quadrupole, which can explain several other CMB anomalies (see \cite{2024JCAP...06..001G,Gaztanaga:2024whs,2836794}).

Our analysis shows that the observed cosmological constant $\Lambda$ can be interpreted as a boundary effect from the gravitational radius of the BHU, aligning with the idea of an effective \(\Lambda\) term without invoking exotic physics. The implications of this model extend to the generation of super-horizon perturbations, the observed entropy ratio of baryons to photons, and the potential origins of dark matter (\cite{gaztanaga:bhu2}).
Future studies should explore the role of temperature and radiation in nuclear fusion during the bounce to provide a more comprehensive understanding of the transition from collapse to expansion. 

{The main limitation of our model is the simplifying assumptions of uniformity and spherical symmetry. However, in contrast to molecular or protostellar clouds, the collapsing fluid in the FLRW cloud is not expected to fragment or form clumps. The gravitational collapse is almost completely  determined by GR and occurs much smoother due to the nearly homogeneous, and cold initial state. Nevertheless, small deviations from uniformity may lead to localized overdensities at later stages.  If these regions exceed a critical threshold before the bounce, they could undergo gravitational collapse and form compact remnants, such as primordial neutron stars or black holes, through purely relativistic gravitational instability. This process is distinct from the cooling-driven fragmentation seen in molecular clouds, because in our scenario there is no radiative cooling mechanism. If sufficiently abundant and long-lived, such objects could naturally provide a non-particle dark matter candidate within the BHU scenario \cite{gaztanaga:bhu}, analogous to but distinct from standard primordial black holes.
It should be emphasized that this possibility remains theoretical and requires further validation. }

{
A realistic treatment calls for a fully relativistic model of gravitational collapse and bounce within standard GR, incorporating realistic equations of state and boundary conditions, as well as numerical simulations to track the evolution of perturbations through the bounce phase. Furthermore, quantum effects are expected to become significant close to the bounce, potentially modifying both the collapse dynamics and the fate of compact remnants. Exploring these quantum effects represents an important avenue for future investigation.}

The smoking gun for our bouncing scenario is the presence of both a small spatial curvature and a small $\Lambda$ term. While the latter has already been measured with high precision, the former remains a testable prediction (given here in Eq.~\ref{eq:omegak}) for upcoming cosmological surveys. The Planck PR3 lensed power spectrum revealed a $3\sigma$ preference for positive curvature \cite{2020A&A...641A...6P}, with $\Omega_k \simeq -0.04 \pm 0.01$, in agreement with our Eq.~\ref{eq:omegak}. Recent results from ACT \cite{2025arXiv250314452L} similarly suggest a slight preference for positive curvature (see their Fig.~9), although the current uncertainties remain too large to decisively rule out a flat universe. The latest DESI data \cite{2025arXiv250314738D} echo this trend, also hinting at a mild preference for positive curvature.
Together, the ACT and DESI results support a growing pattern: when multiple high-precision datasets are combined, persistent tensions with the standard $\Lambda$CDM model begin to emerge. Notably, the combination of DESI and CMB data reveals $3\sigma$ evidence for a $\Lambda$ term that evolves (reduces) slowly over cosmological time. In our framework, where $\Lambda = 3 / r_{\rm S}^2$, this corresponds to an increasing
 FLRW cloud mass $m$ over time. If confirmed, this could be due to mass accretion into the BHU. On the contrary, 
a decreasing FLRW cloud mass could be interpreted as a signature of quantum horizon effects—such as black hole evaporation via Hawking radiation \cite{1975CMaPh..43..199H,Kumar:2023ctp,Kumar:2023hbj}.
Nonetheless, individual cosmological measurements have not yet yielded definitive evidence for departures from the standard $\Lambda$CDM scenario.

\begin{acknowledgments}
{We are grateful to the referee for the valuable feedback, which has significantly contributed to making the presentation of our results clearer and more comprehensive.}
EG acknowledges grants from Spain Plan Nacional (PGC2018-102021-B-100) and Maria de Maeztu (CEX2020-001058-M). KSK acknowledges the support from the Royal Society through the Newton International Fellowship. 
SP acknowledges the support and hospitality at the Institute of Cosmology and Gravitation during a 3-month stay to conclude his Master's Thesis.
MG acknowledges the support through the Generalitat Valenciana via the grant CIDEGENT/2019/031, 
the grant PID2021-127495NB-I00 funded by MCIN/AEI/10.13039/501100011033 and by the European Union, and the Astrophysics and High Energy Physics program of the Generalitat Valenciana ASFAE/2022/026 funded by MCIN and the European Union NextGenerationEU (PRTR-C17.I1).

\end{acknowledgments}

\bibliographystyle{apsrev4-2}
\bibliography{BounceInflation}

\appendix

\section{Appendix A: Analogy with Scalar Field}
\label{Appendix:A}

We can model the quantum ground state of nuclear saturation by introducing a scalar degree of freedom $\psi=\psi(\boldsymbol{x}_\alpha)$.
Consider the Einstein-Hilbert action with minimally coupled matter fields with Lagrangian ${\cal L}$:
\beq
S = \int_{\calMa} d\calMa \left[ \frac{ R}{16\pi G} +  {\cal L} \right]
 \,,
\label{eq:actionC}
\eeq
The energy-momentum $T_{\mu\nu}$ is defined as:
\beq
\boldsymbol{T}_{\mu\nu} \equiv -\frac{2}{\sqrt{g}} \frac{\delta (\sqrt{-g}  {\cal L})}{\delta \boldsymbol{g}^{\mu\nu} }
= \boldsymbol{g}_{\mu\nu} {\cal L}  - 2 \frac{\partial {\cal L}}{\partial \boldsymbol{g}^{\mu\nu}} \,.
\label{eq:TmunuL}
\eeq
The least action principle with respect to the metric $\boldsymbol{g}_{\mu\nu}$
yields  Einstein field equations:
\beq
\frac{\delta S}{\delta \boldsymbol{g}^{\mu\nu} }=0 ~\rightarrow  ~
\boldsymbol{G}_{\mu\nu} \equiv \boldsymbol{R}_{\mu\nu} - \frac{1}{2} \boldsymbol{g}_{\mu\nu} R   = 8 \pi G \boldsymbol{T}_{\mu\nu} \,,
\eeq
For the Lagrangian, we consider a combination of a perfect fluid and an effective minimally coupled scalar field $\psi=\psi(x_\alpha)$ with:
$ {\cal L} =  {\cal L}_m + {\cal L}_\psi$, where
$ {\cal L}_\psi = {1\over{2}} \bar\nabla^2 \psi - V(\psi)$ and ${\cal L}_m$ is the standard matter-energy content.
We have defined $\bar\nabla^2 \psi \equiv \partial_\alpha \psi \partial^\alpha \psi$ and $V(\psi)$ is the potential of the classical scalar field $\psi$. We will next explore the regime where  ${\cal L}$ is dominated by ${\cal L}_\psi$. If both ${\cal L}_m$ and ${\cal L}_\psi$ 
contributions are not coupled, then the general result would correspond to just adding both contributions to $P$ and $\rho$.
We can estimate the contribution of the scalar field $\boldsymbol{T}_{\mu\nu}(\psi)$ from equation\,\eqref{eq:TmunuL}:
\beq
\boldsymbol{T}_{\mu\nu}(\psi) =  \partial_\mu \psi \partial_\nu \psi - \boldsymbol{g}_{\mu\nu} \left[ {1\over{2}}\bar\nabla^2 \psi - V(\psi) \right] \,.
\label{eq:TmunuPsi}
\eeq
Choosing an observer that is moving with the fluid and comparing it to a perfect fluid, we can identify (see also equations B66-B68 in  \cite{Cosmology_book_Weinberg}):
\beq
\rho =    {1\over{2}} \dot\psi^2 + V(\psi)
\quad ; \quad
P  =  {1\over{2}}  \dot\psi^2  - V(\psi) \,, 
\label{eq:relat2}
\eeq
where we have defined $\dot\psi \equiv \partial_0 \psi$.
The ground state $\Psi_{\rm G}$ of the system corresponds to the configuration in which the energy is minimized. In a relativistic context, this often means that the kinetic contributions (derived from the gradient terms \(\partial_\mu \psi\)) become insignificant compared to the potential energy $V_{\rm G} \equiv V(\psi_{\rm G}) \gg \dot\psi^2_{\rm G}$.
This simply means that the total energy is dominated by the potential energy. 
We expect something similar to happen when the collapsing cloud reaches the ground state at some supra-nuclear densities in a cold collapse. The dynamics will be dominated by the potential of the interaction of quantum particles, and their kinetic energy will not play a significant role in the evolution at the bounce. Close to the ground state, $V<V_{\rm G}$. During its evolution, first, the cloud ascends (rolls up) towards the potential $V_{\rm G}$ as it collapses, and after the bounce, it descends (rolls down). Analogously to the scalar field considered here, a fluid with a given EoS will reach some saturation density $\rho=\rho_{\rm G} = V_{\rm G} = -P_{\rm G} = constant$. Consequently, the EoS plays a role in the scalar potential. 

If we take $n_s = 0.9649\pm 0.0042$ (Planck TT+TE+EE+LowE+Lensing) at $k_\ast = 0.05 \,\text{Mpc}^{-1}$ from the Planck data \cite{Planck:2018jri})  we find (considering the spectral index derived for Starobinsky-like inflationary scenario \cite{Kehagias:2013mya}:
\beq
N_e=\frac{2}{2\epsilon+\eta} = \frac{2}{(1-n_s)}  \simeq 56.98^{+7.74}_{-6.09} \,.
\label{eq:Ne}
\eeq
where $\epsilon = -\frac{\dot H}{H^2},\, \eta = \frac{\dot{\epsilon}}{H\epsilon}$ are the slow-roll parameters that characterize the inflationary expansion after the bounce.

These are related to energy density $\rho$ and pressure $P=P(\rho)$ EoS as (in the units of $c=1$)
\begin{equation}
    \epsilon = \frac{\LF \rho+P \RF}{H^2}, \quad \eta = \frac{\dot\epsilon}{H\epsilon} = \frac{2}{H^2}\LF \rho+P \RF -3\LF 1+\frac{\dot P}{\dot \rho} \RF
\end{equation}
which are small during the inflationary expansion, which is followed by (quasi-de Sitter) bounce that occurs in our model due to negative degeneracy pressure. In the case of Starobinsky or Higgs inflation $\epsilon \approx \frac{3}{4N_e^2},\,\eta \approx 2/N_e$. 

The primordial power spectrum (of curvature perturbation $\zeta$) in the framework of single-field slow-roll inflationary models is
\begin{equation}
    P_\zeta = A_s\LF \frac{k}{k_s}\RF^{n_s-1},\quad A_s = \frac{H_{\rm inf}^2}{m_{\rm P}^28\pi^2\epsilon}\Bigg\vert_{k=k_s}
\end{equation}
where $k_s= 0.05\, {\rm Mpc}^{-1}$ is the pivot scale chosen by the Planck data, $A_s$ is the amplitude of the power spectrum, which at the pivot scale measured to be $A_s\sim 2.2\times 10^{-9}$, $H_{\rm inf}$ is the value of Hubble parameter during inflation and $\epsilon = -\frac{\dot{H}_{\rm inf}}{H_{\rm inf}^2}\ll 1$ is the known as slow-roll parameter that measures how slowly Hubble parameter varies during inflation. Using the observational constraint on the amplitude of the power spectrum, the Hubble parameter during inflation can be estimated to be 
\begin{equation}
    H_{\rm inf} = \sqrt{8\pi^2\epsilon} \sqrt{2.2\times 10^{-9}} m_{\rm P} \approx 4.17\times 10^{-4}\sqrt{\epsilon}\, m_{\rm P}
\end{equation}
In the case of Starobinsky inflation $\epsilon = \frac{3}{4N_e^2}$ which yields $H_{\rm inf}\approx 6.33^{+0.76}_{-0.76}\times 10^{-6}\,m_{\rm P}$. 

Although the above measurements are much smaller than the Planck values, they remain significantly larger than the nuclear saturation scale given in Eq.~\ref{eq:NS}. This suggests the existence of a higher-energy ground state beyond neutral saturation. In particular, this energy scale is comparable to those typically considered in inflationary models, such as Starobinsky-like inflationary models \cite{Kehagias:2013mya}. Moreover, it aligns with the observed normalization of the CMB temperature power spectrum, \( A_s \), at the pivot scale \( k_\ast = 0.05 \,\text{Mpc}^{-1} \), reinforcing its relevance in early-universe physics.

\section{Appendix B: Relativistic EoS in the rest frame}
\label{Appendix:B}

Consider a change in variables from comoving coordinates $\boldsymbol{x}^\nu=[\tau,\chi]$ to rest frame coordinates $\boldsymbol{\xi}^\alpha= [t, r]$, where $r=a(\tau) \chi$ is the rest frame radial coordinate, while angular variables $(\theta, \delta)$ remain the same. The most general form for a metric with spherical symmetry can be written in terms of Bardeen potentials  $\Psi(t,r)$ and $\Phi(t,r)$ as:
\beq
ds^2 = \boldsymbol{f}_{\alpha\beta} d\boldsymbol{\xi}^\alpha d\boldsymbol{\xi}^\beta = -({1+2\Psi}) dt^2 + \frac{dr^2}{1+2\Phi} + r^2 d\Omega^2
\label{eq:Bardeen}
\eeq
The FLRW comoving metric $\boldsymbol{g}_{\mu\nu}$ in Eq.\ref{eq:FLRWk} transforms into the rest frame FLRW metric in Eq.\ref{eq:Bardeen} as:
\bea
\boldsymbol{g}_{\mu\nu} &=& \Lambda_\mu^\alpha \Lambda_\nu^\beta  \boldsymbol{f}_{\alpha\beta}, \\
\Lambda_\mu^\alpha \equiv
\frac{\partial \boldsymbol{\xi}^\alpha}{\partial \boldsymbol{x}^\mu} 
&=&  \begin{pmatrix}
  \partial_\tau t &  \partial_\chi t \\
  \partial_\tau r    &  \partial_\chi r  \\
 \end{pmatrix}
\eea
where the angular part is the identity matrix.
Explicitly: 
\beq
\Lambda^T
\begin{pmatrix} 
-(1+2\Psi)
& 0 
 \\ 
0 & (1+2\Phi)^{-1} 
\end{pmatrix}
\Lambda 
=
\begin{pmatrix} 
-1 & 0 
 \\ 
0 & \bar{a}^2 
\end{pmatrix}
\label{eq:fab}
\eeq
where $\bar{a}^2 \equiv a^2/(1-k \chi^2)$. The general solution to these equations is:
\bea
& \Lambda =
\begin{pmatrix} 
(1+2\Phi_{\rm W})^{-1} & \bar{a} r H (1+2\Phi_{\rm W})^{-1} &
 \\ 
r H  &  \bar{a}
\end{pmatrix}  \label{eq:xi2xH} \\
& (1+2\Phi_{\rm W})^2 \equiv (1+2\Psi)(1+2\Phi) & 
\eea
where $\Phi_{\rm W}=\Phi_{\rm W}(t,r)$ is the Weyl potential, where $2\Psi$ is arbitrary and $2\Phi= r^2 H^2$ with $H=H(\tau)=H(t,r)$.
This frame duality can be interpreted as a Lorentz contraction $\gamma=1/\sqrt{1-u^2}$ where the velocity $u$ is given by the Hubble-Lemaitre law: $u= Hr = u(t,r)$. An observer in the rest frame, not moving with the fluid, sees the moving fluid element $a d\chi$ contracted by the Lorentz factor $\gamma$:  $a d\chi = \gamma dr = dr/\sqrt{1-r^2H^2}$ \cite{decceleration}.

We can now use the inverse transformation, $\bar{\Lambda} \equiv \Lambda^{-1}$, in Eq.~\ref{eq:xi2xH} to determine the energy-momentum tensor $\boldsymbol{\bar{T}}_{\alpha}^\beta$ in the rest frame:
\bea
 \boldsymbol{\bar{T}}_{\alpha}^{\beta} &=&  \boldsymbol{f}^{\gamma\beta} \boldsymbol{\bar{T}}_{\alpha\gamma}
 = \boldsymbol{f}^{\gamma\beta} \bar{\Lambda}_\alpha^\mu \bar{\Lambda}^\nu_\gamma \boldsymbol{g}_{\nu\sigma} \boldsymbol{T}_{\mu}^\sigma \nonumber
\\
&=&\frac{1}{1-u^2}
\begin{pmatrix}
-\rho - u^2 P &  \frac{u (\rho+P)}{(1+2\Phi_{\rm W})}
 \\ 
-\frac{u (\rho+P)}{(1+2\Phi_{\rm W})} & P + u^2 \rho \\
\end{pmatrix} 
\eea
We observe that the off-diagonal terms are generally nonzero, indicating that the fluid is moving in the rest frame ($u \neq 0$), as expected. Furthermore, the stress-energy tensor $\boldsymbol{\bar{T}}_{\alpha}^{\beta}$ is no longer uniform, even when $P$ and $\rho$ are, implying that both the rest-frame relativistic pressure $\bar{P}$ and energy density $\bar{\rho}$ depend on time and radius, $(t, r)$.  
At the center ($r = 0$), the energy density and pressure are the same in both frames, i.e., $\bar{\rho} = \rho$ and $\bar{P} = P$. Neglecting the off-diagonal terms—corresponding to large values of $\Psi$, small $u$, or the near-degenerate case $P \approx -\rho$—we can explicitly express the radial dependence in terms of $u = u(t,r) = r H(t,r)$:
\beq
\bar{\rho} = \frac{\rho + u^2 P}{1 - u^2}, 
\quad
\bar{P} = \frac{P + u^2 \rho}{1 - u^2}.
\eeq
In the limit of degenerate pressure (the ground state), where $P = -\rho$, the energy density and pressure remain unchanged in both frames, i.e., $\bar{\rho} = \rho$ and $\bar{P} = P$. 

We conclude that the EoS used in the conventional Newtonian approach to stationary NS has little in common with the EoS in the exact analytical collapsing/expanding uniform solutions in GR. Despite this, \cite{Pradhan} found that numerical solutions for the Newtonian spherical collapse with the 
 EoS parameters inspired by typical NS conditions exhibit remarkable similarities to the exact GR problem presented here when mapped to the equivalent uniform GR problem in the comoving frame.

\end{document}